\documentclass[11pt]{article}
\usepackage[T1]{fontenc}
\usepackage{lmodern}
\usepackage[margin=1in]{geometry}
\usepackage{amsmath,amssymb,amsthm,mathtools}
\usepackage{booktabs,array,enumitem,microtype}
\usepackage[colorlinks=true,linkcolor=blue!50!black,citecolor=blue!50!black,urlcolor=blue!50!black]{hyperref}
\usepackage{xcolor}
\newtheorem{theorem}{Theorem}[section]
\newtheorem{lemma}[theorem]{Lemma}
\newtheorem{corollary}[theorem]{Corollary}
\theoremstyle{definition}
\newtheorem{definition}[theorem]{Definition}

\DeclareMathOperator{\br}{br}
\DeclareMathOperator{\rank}{rank}
\DeclareMathOperator{\supp}{supp}
\DeclareMathOperator{\dist}{dist}
\DeclareMathOperator{\spanop}{span}
\newcommand{\eps}{\epsilon}

\newcommand{\E}{\mathbb E}
\newcommand{\calB}{\mathcal B}
\newcommand{\calD}{\mathcal D}

\setlist{itemsep=3pt,topsep=5pt}
\title{Nonadaptive One-Sided Testing of Binary Rank}
\author{{\bf Nader H. Bshouty}\\ Faculty of Computer Science\\  Technion Israel Institute of Technology\\ bshouty@cs.technion.ac.il}
\date{}

\begin{document}
\maketitle
\begin{abstract}
The binary rank of a binary matrix is the minimum number of disjoint
all-one rectangles that partition its one-entries. We study entry-query
testing of binary rank at most $d$, with one-sided error and nonadaptive queries.
We give a tester using
$\tilde O(d^4/\eps^2)$ queries, independent of the matrix dimensions and a general lower
bound of $\tilde \Omega(d^{3/2}/\eps)$. 
\end{abstract}

\section{Introduction}
Suppose a binary matrix records which people have which attributes. A set of
people who all share a set of attributes gives an all-one rectangle in the
matrix. We ask whether all the one-entries can be divided into a small number
of such rectangles, with each one-entry belonging to exactly one rectangle.
The minimum number needed is called the \emph{binary rank}.

If the matrix is very large, even reading all its entries can be expensive.
Property testing asks a less demanding question: does the matrix have binary
rank at most a given integer $d$, or is it far from every such matrix?
Here ``far'' means that at least an $\eps$ fraction of all entries must change.
The tester may behave arbitrarily on matrices that lie between these two cases.
A \emph{query} asks for the value at one chosen row and column.
Our goal is to use few queries, independently of the matrix dimensions.

We impose two requirements. The tester is \emph{nonadaptive}: it chooses all
queried positions before seeing any entries. It also has \emph{one-sided error}:
it always accepts a matrix of binary rank at most $d$. On every $\eps$-far
matrix, it must reject with probability at least $2/3$. Probability refers to
the tester's random choices. Nonadaptivity allows all entries to be requested
at once. One-sided error gives rejection a useful meaning: the entries already
seen rule out every matrix of binary rank at most $d$.

\subsection{Main results}
We prove the following two results.
\begin{theorem}[Nonadaptive upper bound]\label{thm:upper}
For every integer $d\ge1$ and every $0<\eps<1$, there is a randomized
nonadaptive algorithm that, given entry-query access to a binary matrix
$M\in\{0,1\}^{n\times m}$, makes
\[
O\!\left(
\frac{d^2(d+\log(1/\eps))^2}{\eps^2}
\right)=\tilde O\left(\frac{d^4}{\epsilon^2}\right).
\]
queries and satisfies the following guarantees:
\begin{enumerate}
\item If $M$ has binary rank at most $d$, the algorithm accepts with
probability one.
\item If changing $M$ into a matrix of binary rank at most $d$ requires
changing at least $\eps nm$ entries, the algorithm rejects with
probability at least $2/3$.
\end{enumerate}
\end{theorem}

\begin{theorem}[Nonadaptive lower bound]\label{thm:lower}
Let $d\ge4$ be an integer, let $0<\eps\le1/16$, and let $N$ be an integer
satisfying
$N\ge\ {1024d}/{\eps}$.
Every randomized nonadaptive algorithm with entry-query access to
matrices in $\{0,1\}^{N\times N}$ that
\begin{enumerate}
\item accepts every matrix of binary rank at most $d$ with probability
one, and
\item rejects, with probability at least $2/3$, every matrix for which at least
$\eps N^2$ entries must be changed to obtain a matrix of binary rank
at most $d$.
\end{enumerate}
must make at least
\[
\Omega\left(\frac{d^{3/2}\log(1/\eps)}{\eps}\right)
\]
queries in the worst case.
\end{theorem}
\subsection{Previous work and our contribution}
Binary rank has both an algebraic description, by binary matrix factors,
and a combinatorial description, by disjoint rectangles. We explain their
equivalence in Section~\ref{sec:background}. Parnas's survey~\cite{ParnasSurvey26}
provides broader background on binary rank and related notions of rank.

Earlier general testing results already imply dimension-independent tests.
Alon, Fischer, and Newman~\cite{AFN07} studied testing bipartite graphs with
forbidden induced subgraphs. Parnas, Ron, and Shraibman~\cite[Section 4]{PRS21}
explain how that framework applies to binary rank, although the resulting
dependence on $d$ is very large. Nakar and Ron~\cite{NR18} developed a general
framework for graph partition properties. Its specialization to binary rank
gives a nonadaptive one-sided bound
$\widetilde O(2^{4d}/\eps^4)$.

More direct arguments gave better bounds. Parnas, Ron, and Shraibman~\cite{PRS21}
proved a nonadaptive one-sided bound $O(2^{2d}/\eps^2)$ and an adaptive
one-sided bound $O(2^{2d}/\eps)$. Bshouty~\cite{Bshouty23} improved these to
$\widetilde O(2^d/\eps^2)$ and
$\widetilde O(2^d/\eps)$, respectively. That paper also considers
rectangles with bounded overlap. Here the rectangles are disjoint, which is
the case $s=1$ in its terminology.

Parnas~\cite{Parnas26} recently obtained an adaptive two-sided tester using
$\tilde O(d^3/\eps^2)$ queries. Thus polynomial query complexity itself is
already known. Our upper bound obtains polynomial complexity with both
nonadaptivity and one-sided error. It has a larger dependence on $d$ at
constant $\eps$ than that adaptive two-sided bound.

\begin{table}[ht]
\centering\small
\renewcommand{\arraystretch}{1.35}
\begin{tabular}{@{}lll@{}}
\toprule
Upper-bound source & Model & Queries\\
\midrule
Nakar--Ron~\cite{NR18,PRS21} & Nonadaptive, one-sided &
$\widetilde O(2^{4d}/\eps^4)$\\
Parnas--Ron--Shraibman~\cite{PRS21} & Nonadaptive, one-sided &
$O(2^{2d}/\eps^2)$\\
Bshouty~\cite{Bshouty23} & Nonadaptive, one-sided &
$\widetilde O(2^d/\eps^2)$\\
Parnas~\cite{Parnas26} & Adaptive, two-sided &
$\widetilde O(d^3/\eps^2)$\\
Theorem~\ref{thm:upper} & Nonadaptive, one-sided &
$\widetilde O(d^4/\eps^2)$\\
\bottomrule
\end{tabular}
\caption{Selected upper bounds for binary rank at most $d$.
The models differ as indicated.}
\end{table}

\subsection{The Upper Bound}
One ingredient in our upper-bound proof is also known: the description of
binary factors by cliques and independent sets in a graph, given in
Parnas's work~\cite[Lemma 9]{Parnas26}. We explain below how we use this
construction.

Let $M\in\{0,1\}^{n\times m}$ be the matrix being tested. Our tester
independently samples $t=\tilde O(d^2/\epsilon)$ row indices and $t$ column indices, each
uniformly from the corresponding set of indices. It queries all entries
of the resulting submatrix $S$. It accepts if $S$ has binary rank at
most $d$, and rejects otherwise. Thus, all queries are chosen before
any entry is read. The tester has one-sided error because every submatrix of a matrix
of binary rank at most $d$ also has binary rank at most $d$.

If $S$ is accepted, it has binary rank at most $d$. Therefore, we can
write $S=AB$, where $A\in\{0,1\}^{t\times d}$ and
$B\in\{0,1\}^{d\times t}$, and multiplication uses ordinary arithmetic. In the proof, we select ${\it basis}$ of the rows of $A$. That is, 
linearly independent rows of $A$ that span its row space over
$\mathbb R$. We also select basis of the columns of $B$. Each selection contains at
most $d$ vectors.

We also form a graph with vertices $1,\ldots,d$. Two vertices are joined
by an edge if some row of $A$ has a $1$ in both corresponding positions.
No column of $B$ can have a $1$ in both positions of an edge, because
its product with that row would then be at least $2$. Consequently,
the positions containing~$1$ in each row of $A$ form a clique, and
those in each column of $B$ form an independent set.

For the proof, we also construct a matrix $H\in \{0,1\}^{n\times m}$ that agrees with $M$ at every sampled entry. Write $S_{ij}=M_{r_i,c_j}$, where $r_1,\ldots,r_t$ and
$c_1,\ldots,c_t$ are the sampled row and column indices.
To construct $H$, we use the selected basis rows of $A$, the selected
basis columns of $B$, their positions within $A$ and $B$, and the graph.
For each selected basis row $i$ of $A$, we also use all entries of
row $r_i$ of $M$. For each selected basis column $j$ of $B$, we use
all entries of column $c_j$ of $M$. The construction described below
uses these vectors, positions, graph, and entries to define a matrix
$H\in\{0,1\}^{n\times m}$ of binary rank at most $d$.\footnote{These
full rows and columns of $M$ are used only to define $H$ in the proof.
The tester queries only the entries in $S$.}
We show that whenever $S$ has binary rank at most $d$, the bases and
graph obtained from its factorization $S=AB$ give a matrix $H$ satisfying
\[
H_{r_i,c_j}=M_{r_i,c_j}=S_{ij}
\qquad\text{for all }1\le i,j\le t.
\]
Thus, $H$ agrees with $M$ at every sampled entry.

For the probability analysis, fix a matrix $M$ that is $\epsilon$-far
from every binary matrix of binary rank at most $d$.
We consider all possible choices of the basis rows of $A$, the basis
columns of $B$, their row and column numbers, and the graph.
We examine each such choice separately. If row $i$ of $A$ is selected
for its basis, we hold fixed the index $r_i$. Similarly, if column $j$ of $B$ is selected
for its basis, we hold fixed $c_j$. With these indices fixed, the construction
defines a fixed matrix~$H$. The other sampled indices remain
independent and uniform, since all indices were sampled independently.

Since $H$ has binary rank at most $d$, it differs from $M$ on at least
$\eps nm$ entries. We bound the probability that $H$ and $M$ agree at every entry
determined by a remaining sampled row and a remaining sampled column.
We then bound the number of possible choices of bases, their row and
column numbers, and the graph, and apply a union bound over all these
choices. This bounds the probability of acceptance, because
acceptance implies that at least one choice produces a matrix $H$
that agrees with every queried entry.

\subsection{The Lower Bound}
Our lower-bound proof uses the fact that a one-sided tester can reject
only when the entries it has queried rule out every binary matrix of
binary rank at most $d$. If we can fill in the unqueried entries to
obtain such a matrix while preserving all the answers already given,
then the tester must accept. We call such a filling a completion of
the queried entries.

We construct random binary matrices that are $\eps$-far from every
binary matrix of binary rank at most $d$. We then show that, when the
tester makes too few queries, the entries it observes often have a
completion of binary rank at most $d$. Thus, although the entire matrix
is far from the property, the queried entries do not provide enough
information for a one-sided tester to reject it.

To obtain the dependence on $\eps$, we place random bits in submatrices
with different combinations of numbers of rows and columns. The proof controls both the distance of the entire
matrix from the property and the probability that the queried entries
still admit a completion of binary rank at most $d$. Together, these
estimates give the explicit lower bound.

\section{Background and notation}\label{sec:background}
\subsection{Rectangles and binary factors}
For a positive integer $a$, write $[a]=\{1,\ldots,a\}$. A binary matrix has
entries in $\{0,1\}$. A rectangle is a set $R\times C$ of matrix positions,
where $R$ is a set of row indices and $C$ is a set of column indices.

\begin{definition}
The \emph{binary rank} $\br(M)$ of a binary matrix $M$ is the smallest number
of all-one rectangles whose disjoint union is the set of one-entries of $M$.
The all-zero matrix has binary rank zero. We write $\calB_d^{n,m}$ for the set
of $n\times m$ binary matrices with binary rank at most $d$.
\end{definition}

Matrix multiplication gives a useful way to record the rectangles. Assign
one coordinate to each rectangle. Put $A_{iu}=1$ when row index $i$ belongs to
rectangle $u$, and put $B_{uj}=1$ when column index $j$ belongs to it. Then
\[
(AB)_{ij}=\sum_{u=1}^d A_{iu}B_{uj}
\]
counts the rectangles containing position $(i,j)$. Therefore $\br(M)\le d$
is equivalent to the existence of a factorization
\[
M=AB,\qquad A\in\{0,1\}^{n\times d},\quad
B\in\{0,1\}^{d\times m},
\]
with ordinary integer multiplication. Conversely, each column $u$ of $A$ and row $u$ of $B$ define the
rectangle $\{i:A_{iu}=1\}\times\{j:B_{uj}=1\}$. Since each entry is zero or one, the
rectangles cannot overlap. If fewer than $d$ rectangles are needed, append
zero columns to $A$ and zero rows to $B$.

This differs from \emph{Boolean rank}, where sums are replaced by
logical OR, and the rectangles covering the entries equal to $1$
may overlap. All linear algebra in this paper is over the real numbers.

The \emph{support} of a binary vector $a$, denoted by $\supp(a)$,
is the set of indices at which its entries equal $1$.
Thus, for two binary vectors $a$ and $b$ of the same length,
their inner product $a\cdot b$ equals
$|\supp(a)\cap\supp(b)|$.
We will also use the fact that $\rank(M)\le\br(M)$, whereas the
reverse inequality need not hold.

\subsection{Distance, partial answers, and one-sided error}
For $M\in \{0,1\}^{n\times m}$, the {\it normalized
Hamming distance} from a matrix $H\in \{0,1\}^{n\times m}$ is
\[
\dist(M,H)=\frac{|\{(i,j):M_{ij}\ne H_{ij}\}|}{nm}.
\]
The matrix $M$ is \emph{$\eps$-far} from $\calB_d^{n,m}$ when this distance
is at least $\eps$ for every $H\in\calB_d^{n,m}$. 

A \emph{transcript} records each position queried by the tester
and the value of the matrix entry at that position.
We view it as a partial matrix: queried entries are specified and other
entries are blank. A \emph{completion} fills the blanks without changing
specified entries. For example, the partial matrix
\[
\begin{pmatrix}1&*\\ *&1\end{pmatrix}
\]
has a binary-rank-one completion, obtained by filling both blanks with ones.
But specifying its upper-right entry as zero rules out rank one.

This connection between partial answers and completions is the basis of the
lower bound.
\begin{lemma}[One-sided rejection requires an obstruction]\label{lem:transcript}
Fix a seed for which the resulting deterministic nonadaptive tester
accepts every matrix in $\calB_d^{n,m}$. It can reject a transcript
only if that transcript has no completion in $\calB_d^{n,m}$.
\end{lemma}
\begin{proof}
Suppose there is a matrix $H\in\calB_d^{n,m}$ that agrees with
all the queried entries. The tester sees exactly the same answers
on $H$, because its query positions were chosen before any answers.
Since it accepts $H$, it must accept this transcript.
\end{proof}

\subsection{Graphs and elementary probability}
A \emph{proper coloring} of a graph assigns colors to vertices
so that endpoints of an edge have different colors. Each set of vertices
with one color is therefore independent. 
We use cliques and independent sets to describe the rows of $A$
and the columns of $B$ in a binary factorization $M=AB$.
We use graph colorings to construct completions of partial matrices.

We need a bound for a sum of independent zero-one variables. 
\begin{lemma}[Chernoff bound]\label{lem:chernoff}
Let $Z$ be a sum of independent random variables taking values zero and one,
and let $\mu=\mathbb E [Z]$. For $0<\delta<1$,
\[
\Pr[Z\le(1-\delta)\mu]\le \exp(-\delta^2\mu/2).
\]
\end{lemma}

\section{A general polynomial upper bound}
\subsection{Globally valid reconstructions}
Our tester accepts when the sampled submatrix has binary rank at most $d$.
To analyze acceptance on an $\eps$-far input $M$, we construct a binary
matrix $H$ with the same dimensions as $M$ and binary rank at most $d$.
The matrix $H$ will agree with $M$ at every sampled entry.

We construct $H$ as the product of two binary matrices, the first with
$d$ columns and the second with $d$ rows. However, such a product need
not be binary. If a row of the first matrix and a column of the second
both have $1$ in two common positions, the corresponding entry of their
product is at least $2$. The following known graph construction ensures
that every entry of the product is $0$ or $1$. We include its short
proof because we use this construction to define $H$.
\begin{lemma}[Parnas~{\cite[Lemma 9]{Parnas26}}]\label{lem:graph}
Let $A\in\{0,1\}^{n\times d}$ and $B\in\{0,1\}^{d\times m}$,
and let $G$ be a simple graph on $[d]$. Suppose the support\footnote{Recall that, the support of a vector $v=(v_1,\ldots,v_d)$ is the set
$\{i\in[d]:v_i\ne0\}$. For a binary vector, these are exactly the
indices at which the entries equal $1$.} of
every row of $A$ is a clique in $G$, and the support of every
column of $B$ is an independent set in $G$. Then every entry
of $AB$ belongs to $\{0,1\}$, and $\br(AB)\le d$.

Conversely, if $M=AB$ is a binary matrix, then there exists a
simple graph $G$ on $[d]$ such that the support of every row
of $A$ is a clique in $G$, and the support of every column
of $B$ is an independent set in $G$.
\end{lemma}
\begin{proof}
A clique and an independent set intersect in at most one vertex, so their
binary vectors have inner product at most one. For the converse, put an edge
$\{u,v\}$ in $G$ whenever some row of $A$ contains both coordinates. A column
of $B$ cannot contain both endpoints of such an edge, since its product with
that row would be at least two. 
\end{proof}
We next construct a binary matrix $H$ of the same dimensions as $M$
and of binary rank at most $d$. The construction uses a graph,
two lists of binary vectors, and entries in selected rows and columns
of $M$. We will later show how to choose these so that $H$ agrees
with $M$ on an accepted sample.

Choose lists of row indices $I=(i_1,\ldots,i_k)$ and column indices
$J=(j_1,\ldots,j_\ell)$ of $M$, where $0\le k,\ell\le d$.
Choose binary matrices
$X\in\{0,1\}^{k\times d}$,
and
$Y\in\{0,1\}^{d\times\ell}$,
and a simple graph $G$ on $[d]$. 

Let $U$ be the real linear space spanned by the rows of $X$, and
let $V$ be the real linear space spanned by the columns of $Y$:
$U=\spanop\{\text{rows of }X\}$,
and
$V=\spanop\{\text{columns of }Y\}$.
The span of an empty list is $\{0\}$. We use the notation
\[
M[i,J]=(M_{i,j_1},\ldots,M_{i,j_\ell}),
\qquad
M[I,j]=(M_{i_1,j},\ldots,M_{i_k,j})^{\mathsf T}.
\]
Thus, $M[i,J]$ contains the entries of row $i$ in the selected
columns, and $M[I,j]$ contains the entries of column $j$ in the
selected rows.

For each row $i\in[n]$, consider all binary row vectors $a$ of
length $d$ satisfying
\[
a\in U,\qquad
\supp(a)\text{ is a clique in }G,\qquad
aY=M[i,J].
\]
The last equation requires $a$ to have inner product $M_{i,j_q}$
with column $q$ of $Y$, for every $q\in[\ell]$.
If such vectors exist, let $\widehat a_i$ be the first one in
lexicographic order. 
If no such vector exists, set $\widehat a_i=0$.

Similarly, for each column $j\in[m]$, consider all binary column
vectors $b$ of length $d$ satisfying
\[
b\in V,\qquad
\supp(b)\text{ is an independent set in }G,\qquad
Xb=M[I,j].
\]
The last equation requires $b$ to have inner product $M_{i_p,j}$
with row $p$ of $X$, for every $p\in[k]$.
Let $\widehat b_j$ be the lexicographically first such vector,
or set $\widehat b_j=0$ if none exists.

Define
\begin{equation}\label{eq:decoder}
H_{ij}=\widehat a_i\widehat b_j
\qquad (i\in[n],\ j\in[m]).
\end{equation}
The support of every chosen row vector is a clique, and the support
of every chosen column vector is an independent set. This remains
true when we choose the zero vector, whose support is empty.
By Lemma~\ref{lem:graph}, every entry of $H$ belongs to $\{0,1\}$.
The chosen vectors give a binary factorization of $H$ with $d$
columns in the first matrix and $d$ rows in the second.
Hence $H\in\calB_d^{n,m}$.

For fixed $M,I,J,X,Y,G$, the construction determines $H$ uniquely.
It always gives a matrix of binary rank at most $d$, although $H$
need not agree with $M$ at the selected entries. In particular,
choosing the zero vector when no solution exists need not satisfy
the corresponding equation.

Constructing all the vectors $\widehat a_i$ uses all entries in
the columns listed in $J$. Constructing all the vectors
$\widehat b_j$ uses all entries in the rows listed in $I$.
These entries are used only to define $H$ in the proof.
The tester queries only the sampled submatrix and does not
construct $H$.

The next lemma shows that, whenever the sampled submatrix $S$ has
binary rank at most $d$, we can choose $I,J,X,Y,G$ in the preceding
construction so that $H$ agrees with $M$ at every sampled entry.

\begin{lemma}[Encoding an accepted sample]\label{lem:encoding}
Let $S$ be a sampled $t\times t$ submatrix, where repeated sampled indices
are permitted. If $\br(S)\le d$, some description of the preceding form,
using at most $d$ sampled row positions and $d$ sampled column positions,
produces a matrix $H$ agreeing with every entry of $S$.
\end{lemma}
\begin{proof}
Write $S_{ij}=M_{r_i,c_j}$, where $r_1,\ldots,r_t$ and
$c_1,\ldots,c_t$ are the sampled row and column indices.
Since $\br(S)\le d$, we can write
\[
S=AB,\qquad
A\in\{0,1\}^{t\times d},\qquad
B\in\{0,1\}^{d\times t}.
\]
If necessary, add zero columns to $A$ and zero rows to $B$
to obtain these dimensions. Denote row $i$ of $A$ by $a_i$
and column $j$ of $B$ by $b_j$. Thus, $S_{ij}=a_i b_j$.

Choose linearly independent rows of $A$ that span its row space
over $\mathbb R$. Let their positions in $A$ be $p_1,\ldots,p_k$,
and let $X$ be the matrix consisting of these rows.
Similarly, choose linearly independent columns of $B$ that span
its column space over $\mathbb R$. Let their positions in $B$
be $q_1,\ldots,q_\ell$, and let $Y$ consist of these columns.
Both $k$ and $\ell$ are at most $d$.

In the preceding construction, use
\[
I=(r_{p_1},\ldots,r_{p_k}),
\qquad
J=(c_{q_1},\ldots,c_{q_\ell}).
\]
Choose the graph $G$ supplied by Lemma~\ref{lem:graph} for
the factorization $S=AB$. Thus, the support of every row of $A$
is a clique in $G$, and the support of every column of $B$
is an independent set in $G$.
By the choices of $X$ and $Y$, the spaces $U$ and $V$ in
the construction are the row space of $A$ and the column
space of $B$, respectively.

For each sampled row position $i$, the vector $a_i$ satisfies
all the conditions used to choose $\widehat a_{r_i}$.
Indeed, $a_i\in U$, its support is a clique in $G$, and
\[
a_iY
=(S_{i,q_1},\ldots,S_{i,q_\ell})
=M[r_i,J].
\]
Therefore, at least one vector satisfies these conditions.
The construction chooses such a vector $\widehat a_{r_i}$,
which also satisfies $\widehat a_{r_i}Y=M[r_i,J]$. Subtracting gives
\[
(\widehat a_{r_i}-a_i)Y=0.
\]
Since the columns of $Y$ span $V$, the vector
$\widehat a_{r_i}-a_i$ has inner product zero with every vector
in $V$.

Similarly, for each sampled column position $j$, the vector
$b_j$ belongs to $V$, its support is an independent set in $G$,
and
\[
Xb_j
=(S_{p_1,j},\ldots,S_{p_k,j})^{\mathsf T}
=M[I,c_j].
\]
Thus, the construction chooses a vector $\widehat b_{c_j}\in V$
satisfying the same equation. Consequently,
\[
X(\widehat b_{c_j}-b_j)=0.
\]
Since the rows of $X$ span $U$, every vector in $U$ has inner
product zero with $\widehat b_{c_j}-b_j$.

We can now compare $H$ and $S$. For every $i,j\in[t]$,
\[
\begin{aligned}
H_{r_i,c_j}-S_{ij}
&=\widehat a_{r_i}\widehat b_{c_j}-a_i b_j\\
&=(\widehat a_{r_i}-a_i)\widehat b_{c_j}
  +a_i(\widehat b_{c_j}-b_j)=0.
\end{aligned}
\]
The first term is zero because $\widehat b_{c_j}\in V$.
The second is zero because $a_i\in U$.
Hence $H_{r_i,c_j}=S_{ij}=M_{r_i,c_j}$ at every sampled entry.
\end{proof}

\subsection{A validation grid and the tester}
We now know that an accepted sample has a short description. The next step
is probabilistic: a fixed candidate that differs from the input in many places
should fail on a random grid. 

\begin{lemma}[Avoiding a set of errors]\label{lem:grid}
Suppose two fixed binary matrices differ on at least an $\eps$
fraction of their entries. Sample $a$ row indices and $b$ column
indices independently and uniformly, with replacement. The probability
that the resulting grid contains no disagreement is at most
\[
e^{-\eps\min\{a,b\}}.
\]
\end{lemma}
\begin{proof}
Let $r_1,\ldots,r_a$ and $c_1,\ldots,c_b$ be the sampled row
and column indices, and put $k=\min\{a,b\}$.
The positions
$(r_1,c_1),\ldots,(r_k,c_k)$
are independent and uniformly distributed over all matrix positions.
At each position, the probability of agreement is at most $1-\eps$.
Therefore, the probability of agreement at all these positions
is at most $(1-\eps)^k\le e^{-\epsilon\min\{a,b\}}$.
\end{proof}

We can now analyze the tester. Its rule is simple: sample rows and columns,
read their intersections, and accept exactly when these entries admit a
binary factorization through $d$ coordinates.

\begin{theorem}\label{thm:upper}
For every $d\ge1$ and $0<\eps<1$, there is a nonadaptive one-sided tester
for $\calB_d^{n,m}$ using
\[
O\!\left(\frac{d^2(d+\log(1/\eps))^2}{\eps^2}\right)
\]
entry queries. 
\end{theorem}
\begin{proof}
Choose
\begin{equation}\label{eq:t}
t=\left\lceil\frac{64d(d+\log(1/\epsilon))}{\eps}\right\rceil.
\end{equation}
Independently sample row indices $r_1,\ldots,r_t$ uniformly from
$[n]$ and column indices $c_1,\ldots,c_t$ uniformly from $[m]$,
with replacement. Define
$S_{ij}=M_{r_i,c_j}$ for $i,j\in[t]$.
The tester queries these entries and accepts if and only if
$\br(S)\le d$. If $\br(M)\le d$, then $\br(S)\le d$ for every
sample, so the tester always accepts.

For the remainder of the proof, fix a matrix $M$ that is
$\eps$-far from every matrix in $\calB_d^{n,m}$.
All probabilities below are over the sampled indices.

Let $\mathcal D$ be the set of all choices
$\theta=(P,Q,X,Y,G)$,
where $P,Q\subseteq[t]$, $|P|,|Q|\le d$,
$X\in\{0,1\}^{|P|\times d}$,
$Y\in\{0,1\}^{d\times|Q|}$,
and $G$ is a simple graph on $[d]$.
The sets $P$ and $Q$ specify positions in the sample.
We list their elements in increasing order.
For each $\theta\in\mathcal D$, define
$Z_\theta=
\bigl((r_i)_{i\in P},(c_j)_{j\in Q}\bigr)$.
Thus, $Z_\theta$ records the sampled indices at the selected
positions. Let $H^\theta$ be the matrix obtained from the
construction~\eqref{eq:decoder}, using
$I=(r_i)_{i\in P}$, $J=(c_j)_{j\in Q}$,
together with $X,Y,G$. Define the event
\[
E_\theta=
\left\{
H^\theta_{r_i,c_j}=M_{r_i,c_j}
\text{ for every }i,j\in[t]
\right\}.
\]

Conditioning on any possible value $Z_\theta=z$ makes
$H^\theta$ fixed. The remaining sampled indices are still
independent and uniform. Since $H^\theta\in\calB_d^{n,m}$,
it differs from $M$ on at least an $\eps$ fraction of entries.
Moreover, $E_\theta$ requires agreement on the grid formed by
the remaining $t-|P|$ row samples and $t-|Q|$ column samples.
Both numbers are at least $t-d$.
Lemma~\ref{lem:grid} therefore bounds the conditional probability
of $E_\theta$ by $e^{-\eps(t-d)}$.

We now obtain the probability bound:
\begin{equation}\label{eq:upper-failure}
\begin{aligned}
\Pr_{r,c}[\br(S)\le d]
&\le
\Pr\left[\bigcup_{\theta\in\mathcal D}E_\theta\right]
&&\text{(Lemma~\ref{lem:encoding})}\\
&\le
\sum_{\theta\in\mathcal D}\Pr[E_\theta]\le
|\mathcal D|\max_{\theta\in\mathcal D}\Pr[E_\theta]
&&\text{(union bound)}\\
&=
|\mathcal D|\max_{\theta\in\mathcal D}
\sum_z
\Pr[Z_\theta=z]\Pr[E_\theta\mid Z_\theta=z]
&&\text{(law of total probability)}\\
&\le
|\mathcal D|\max_{\theta\in\mathcal D}
\sum_z
\Pr[Z_\theta=z]e^{-\eps(t-d)}
&&\text{(the conditional bound above)}\\
&=
|\mathcal D|e^{-\eps(t-d)}
&&\text{(the probabilities sum to one).}
\end{aligned}
\end{equation}
Here each sum over $z$ ranges over all possible values of
$Z_\theta$.

It remains to count $\mathcal D$ and substitute the value of $t$.
For $|P|=k$, there are $\binom tk$ choices of $P$ and
$2^{kd}$ choices of $X$. The same count applies to $Q,Y$.
There are $2^{\binom d2}$ choices of $G$. Hence
\[
\begin{aligned}
|\mathcal D|
=
2^{\binom d2}
\left(\sum_{k=0}^d\binom tk\,2^{kd}\right)^2\le
2^{2d^2+\binom d2}
\left(\sum_{k=0}^d\binom tk\right)^2\le
2^{3d^2}(d+1)^2(t+1)^{2d}\le
2^{3d^2}(t+1)^{2d+2},
\end{aligned}
\]
where the last inequality uses $t\ge d$.

Now it is easy to see that
$$|\mathcal D|e^{-\eps(t-d)}\le 2^{3d^2}(t+1)^{2d+2}e^{-\eps(t-d)}<\frac{1}{3}.$$
Thus, the tester accepts an $\eps$-far input with probability
at most $1/3$, and therefore rejects it with probability
at least $2/3$. The number of queries is at most
\[
t^2=
O\!\left(\frac{d^2(d+\log(1/\eps))^2}{\eps^2}\right).
\]
\end{proof}

\section{Completions of partial transcripts}
We turn to the lower bound. By Lemma~\ref{lem:transcript}, a one-sided tester
cannot reject while its answers still admit a completion of binary rank at
most $d$. Our task is therefore to build low-rank completions from sparse
answers. A completion agrees with the input matrix at every queried entry. At unqueried entries, it may differ from the input matrix.

We construct a completion whose one-entries can be partitioned into
a small number of rectangles. Fix any subset of $h$ columns, where
$0\le h\le m$. In each chosen column, fill every unqueried entry
with zero. The ones in that column then form at most one rectangle,
consisting of that column and the rows containing a one.

For the remaining columns, form a graph whose vertices are the rows.
Join two rows by an edge if some remaining column contains a queried
zero in one row and a queried one in the other. Color the vertices
so that the endpoints of every edge have different colors.

Consider all rows of one color. In any remaining column, their
queried entries have the same value, if any have been queried.
Fill the unqueried entries in these rows with that value, or with
zero if none has been queried. These rows now become identical
on the remaining columns, so their one-entries form at most one
rectangle. Rectangles from different colors use disjoint sets of rows.
They use only columns outside the fixed set of $h$ columns,
so they are also disjoint from the rectangles constructed
in those $h$ columns. Thus, the completion has binary rank at most
$h$ plus the number of colors.

By Diestel~\cite[Proposition 5.2.1]{Diestel17}, every graph
with $e$ edges can be colored with at most $1+\sqrt{2e}$ colors.
Applying this bound to the row-conflict graph gives a completion
of binary rank at most $h+1+\sqrt{2e}$. We give the proof for completeness. 

\begin{lemma}[Column removal and row coloring]\label{lem:completion}
Let $M$ be a partial binary matrix with $m$ columns.
Fix any subset of $h$ columns, where $0\le h\le m$.
Define the \emph{row-conflict graph} to have one vertex for each
row of $M$. Two vertices are joined by an edge if some column
outside the fixed subset has a specified zero in one of the
corresponding rows and a specified one in the other.
If this graph has $e$ edges, then $M$ has a completion of binary
rank at most
\[
h+1+\sqrt{2e}.
\]
\end{lemma}
\begin{proof}
Let $k$ be the minimum number of colors needed by the row-conflict
graph. We show that $k\le 1+\sqrt{2e}$. If $k\le1$, the claimed color bound holds. Otherwise, repeatedly
remove vertices while the remaining graph still needs $k$ colors. In the
resulting minimal graph, every vertex has at least $k-1$ neighbors. Indeed,
take any vertex $v$ and suppose that it has fewer than $k-1$
neighbors. By the minimality of the graph, removing $v$ leaves
a graph that can be colored with $k-1$ colors. In such a coloring,
the neighbors of $v$ use at most $k-2$ colors. Therefore, at least
one of the $k-1$ colors is not used by any neighbor of $v$.
Assigning that color to $v$ gives a coloring of the entire graph
with $k-1$ colors, contradicting the assumption that it requires
$k$ colors.

This graph has at least $k$ vertices, since fewer vertices can be colored
with fewer colors. The sum of its vertex degrees is at least $k(k-1)$.
Each edge is counted twice in that sum. The original graph therefore has
$e\ge k(k-1)/2$ edges. Since $(k-1)^2\le k(k-1)$, this gives
$k\le1+\sqrt{2e}$.
\end{proof}

The following corollary gives sufficient bounds on $h$ and $e$
to ensure that this quantity is at most~$d$.

\begin{corollary}\label{cor:threshold}
For $d\ge4$, a transcript is completable in binary rank at most $d$ whenever
$h<d/2$ and $e<d^2/32$ in Lemma~\ref{lem:completion}.
\end{corollary}
\begin{proof}
The completion rank is less than $d/2+1+d/4\le d$.
\end{proof}

As a first application, the number of specified entries alone already
limits how large a rank can be forced. This explains why an exponent $3/2$
will appear in the query lower bound.

\begin{corollary}\label{cor:qcompletion}
Every partial binary matrix with $q\ge1$ specified entries has a completion
of binary rank at most $3q^{2/3}+1$.
\end{corollary}
\begin{proof}
Let $u=\lceil q^{1/3}\rceil$. Select all columns containing
more than $u$ specified entries.
Their number is at most $q/u$. Each other column with $k$ specified entries
contributes at most $\binom k2\le uk/2$ conflict edges. Thus $e\le uq/2$,
and Lemma~\ref{lem:completion} gives a completion of rank at most
$q/u+1+\sqrt{uq}$. Here $u\ge q^{1/3}$ and $u\le2q^{1/3}$ for $q\ge1$,
so the two nonconstant terms are at most $q^{2/3}$ and
$\sqrt2q^{2/3}$, respectively. Their sum is at most $3q^{2/3}$.
\end{proof}

Corollary~\ref{cor:qcompletion} implies that every partial binary
matrix with no completion of binary rank at most $d$ must have
$\Omega(d^{3/2})$ specified entries. To get a dependence on
$\eps$, we need more: 
we place random entries inside a small, randomly chosen
submatrix and set all other entries to zero. We then show that
a tester making too few queries is unlikely to query enough
entries inside this submatrix to rule out a completion of
binary rank at most $d$.

\section{A general nonadaptive lower bound}
\subsection{Far matrices in a hidden rectangle}
The hard inputs have random bits in a small rectangle and zeros elsewhere.
The rectangle must be large enough that these bits cannot be repaired by
changing only an $\eps$ fraction of the whole matrix. At the same time, we
want to vary its shape while keeping its expected area fixed. 

Fix $0<\eps\le1/16$, put $\rho=8\eps$, and take
$\alpha,\beta\in[\rho,1]$ with $\alpha\beta=\rho$. On an $N\times N$ matrix,
select rows independently with probability $\alpha$ and columns
independently with probability $\beta$. Call the resulting rectangle an {\it active} rectangle. Inside the active rectangle put
independent fair bits. Set all other entries to zero. Denote this
distribution by $\calD_{\alpha,\beta}$. An ``active'' row or column simply
means one selected by this random experiment. For example, if $\rho=1/8$,
we may use $(\alpha,\beta)=(1,1/8)$, $(1/2,1/4)$, or $(1/8,1)$.
Each has the same expected active area $N^2/8$, but a different shape.

Why should the resulting matrix be far from small binary rank? A fixed
candidate matrix of binary rank at most $d$ is unlikely to match many independent fair bits. There
are many candidate matrices, but their number grows like $2^{2Nd}$, while
the number of random bits grows like $\eps N^2$. Taking $N$ large enough
makes the latter dominate.

\begin{lemma}\label{lem:far}
Let $\alpha,\beta\in(0,1]$ satisfy $\alpha\beta=\rho:=8\eps$. If $d\ge1$ and $N\ge1024d/\eps$, a matrix from
$\calD_{\alpha,\beta}$ is $\eps$-far from $\calB_d^{N,N}$ with probability
at least $0.99$.
\end{lemma}
\begin{proof}
We first show that the active rectangle is large. We then bound the chance
of being close to one fixed candidate and sum over all candidates.

Let $a,b$ be the numbers of active rows and columns. Their expectations are
$\alpha N$ and $\beta N$. Applying Lemma~\ref{lem:chernoff} with
$\delta=1/4$ to each count and using the union bound gives
\[
\Pr[a<3\alpha N/4\text{ or }b<3\beta N/4]
\le2e^{-\rho N/32}.
\]
Outside this event, the active area $A=ab$ is at least
$9\rho N^2/16=(9/2)\eps N^2$. 

Condition on these active sets. For any fixed binary matrix $H$, each of
the $A$ random bits disagrees with $H$ independently with probability $1/2$.
Thus the number of disagreements has mean $A/2$. Since $\eps N^2\le A/4$,
Lemma~\ref{lem:chernoff} with $\delta=1/2$ bounds the probability of at most
$\eps N^2$ disagreements by $e^{-A/16}\le \exp{(-9\rho N^2/256)}$. Disagreements outside the active
rectangle can only increase the distance, so ignoring them is safe.

There are at most $2^{2Nd}$ matrices in $\calB_d^{N,N}$. A union bound yields failure probability
at most
\[
2e^{-\rho N/32}
+\exp\!\left(2Nd\log2-\frac{9\rho N^2}{256}\right)<0.01
\]
under the stated lower bound on $N$. 
\end{proof}

\subsection{The contribution of one queried column}
Consider a tester whose queries have already been chosen. A column queried
many times might reveal many active rows; we pay for that column by giving
it its own rectangle. In a column queried few times, we instead count pairs
of active queried rows that could create conflict edges. The threshold
below balances these two ways of completing the answers.

Fix a deterministic nonadaptive query that makes at most $q$ queries and let  $Q$  be its set of queried positions. Let $k_j$ count its
distinct queried positions in column $j$. Thus $\sum_j k_j=|Q|\le q$.
Call column $j$ heavy at scale $\alpha$ when $\alpha k_j\ge\sqrt d$.
Here $\alpha k_j$ is its expected number of queried active rows. The label
``heavy'' depends only on the fixed query set and on $\alpha$, not on the
random active sets or their bits.

\begin{lemma}\label{lem:scale}
For $d\ge4$, the rejection probability of any one-sided tester using $Q$
on $\calD_{\alpha,\beta}$ is at most
\begin{equation}\label{eq:scale}
\frac{16\rho}{d^{3/2}}
\sum_{j:k_j>0} k_j f\!\left(\frac{\alpha k_j}{\sqrt d}\right),
\qquad f(x)=\min\{x,x^{-1}\}.
\end{equation}
\end{lemma}
\begin{proof}
We use Corollary~\ref{cor:threshold}: unless there are many active heavy
columns or many conflict edges elsewhere, the transcript has a low-rank
completion. We bound the expectations of these two quantities and then
apply Markov's inequality.

Restrict the observed partial matrix to the active rows and active columns.
Handle its heavy columns separately, and let $h$ be their number. Then
\[
\E [h]=\beta\,|\{j:\alpha k_j\ge\sqrt d\}|.
\]
The formula holds because each heavy column is active with probability
$\beta$. Let $e$ be the number of conflict edges on the remaining columns. A pair
of queried rows is simultaneously active with probability $\alpha^2$.
Its column is active independently with probability $\beta$. There are
at most $k_j^2/2$ pairs in column $j$. Counting an edge more than once, and
ignoring the further requirement that its two bits be opposite, only
increase the bound. Linearity of expectation therefore gives
\[
\E [e]\le\frac{\beta\alpha^2}{2}
\sum_{\alpha k_j<\sqrt d} k_j^2.
\]
Any completion on the active rectangle can be extended by zeros outside
it, matching every outside query without increasing binary rank.
By Lemma~\ref{lem:transcript} and Corollary~\ref{cor:threshold},
rejection therefore requires
$h\ge d/2$ or $e\ge d^2/32$. Markov's inequality gives
\begin{align*}
\Pr[\text{reject}]
&\le \frac{2\beta}{d}|\{j:\alpha k_j\ge\sqrt d\}|
+\frac{16\beta\alpha^2}{d^2}
\sum_{\alpha k_j<\sqrt d}k_j^2\\
&\le\frac{16\beta}{d}
\sum_j\min\{1,\alpha^2k_j^2/d\}.
\end{align*}
For $k_j>0$, put $x=\alpha k_j/\sqrt d$. Since $\min\{1,x^2\}=xf(x)$
and $\alpha\beta=\rho$, each term in the last sum becomes the corresponding
term of~\eqref{eq:scale}. Columns with $k_j=0$ contribute zero.
\end{proof}

\subsection{Summing over aspect ratios}
A single shape gives only part of the lower bound. Nonadaptivity now becomes
crucial: the tester must use the same distribution of query sets no matter
which shape hides the random bits. We use a sequence of shapes in which the
row activation probability halves at each step and the column activation
probability doubles. Their product, and hence the expected area, stays fixed.

\begin{theorem}\label{thm:lower}
Let $d\ge4$ and $0<\eps\le1/16$. For every $N\ge1024d/\eps$, any
randomized nonadaptive one-sided tester for $\calB_d^{N,N}$ that makes
 $q$ queries satisfies
\[
q=\Omega\!\left(\frac{d^{3/2}}{\eps}\log\frac{2}{\eps}\right).
\]
\end{theorem}
\begin{proof}
We apply the tester to several input distributions, each using a different
choice of $\alpha$ and~$\beta$. Since each distribution produces an
$\eps$-far matrix with probability at least $0.99$, the tester must reject
with probability at least $0.99\cdot(2/3)$ under each distribution.
However, Lemma~\ref{lem:scale} bounds the sum of these rejection
probabilities in terms of the number of queries. Comparing the two
bounds gives the required lower bound on the number of queries.

\paragraph{Step 1: choose the shapes.}
Recall that $\rho=8\eps$. Set $T=\lfloor\log_2(1/\rho)\rfloor$ and use the $T+1$ distributions with
\[
\alpha_t=2^{-t},\qquad\beta_t=\rho2^t,
\qquad 0\le t\le T.
\]
All these parameters lie in $[\rho,1]$, and $\alpha_t\beta_t=\rho$,
so Lemmas~\ref{lem:far} and~\ref{lem:scale} apply.

\paragraph{Step 2: each column contributes little in total.}
Recall that $f(x)=\min\{x,x^{-1}\}$. For any $x>0$, the sum $\sum_{t=0}^T f(2^{-t}x)$ is at most $4$.
To verify this, begin with the first argument at most one. Its value is
at most one, and subsequent values halve, so their sum is at most
$1+1/2+1/4+\cdots=2$. For the arguments greater than one, read the sequence
in the reverse order. Their reciprocals again start below one and halve,
so their sum is at most two. Either part may be empty.

Summing Lemma~\ref{lem:scale} over these distributions for a fixed query
set gives
\[
\sum_{t=0}^T\Pr_{\calD_{\alpha_t,\beta_t}}[\text{reject}]\le \frac{16\rho}{d^{3/2}}\sum_{j:k_j>0} k_j \sum_{t=0}^Tf\!\left(\frac{2^{-t} k_j}{\sqrt d}\right)
\le \frac{64\rho q}{d^{3/2}}.
\]
For a randomized tester, this estimate is first applied to every
fixed random seed, using
Lemma~\ref{lem:transcript}, and then averaged over seeds. For each seed,
nonadaptivity fixes the same query set for every shape, and the worst-case
budget guarantees $|Q|\le q$.

\paragraph{Step 3: soundness forces many queries.}
By Lemma~\ref{lem:far} and soundness, the tester rejects with probability
at least $0.99(2/3)>3/5$ under each of these unconditioned distributions.
Consequently,
\[
\frac35(T+1)\le\frac{64\rho q}{d^{3/2}}.
\]
It remains to compare the number of shapes with the desired logarithm.
Write $u=\log_2(1/\rho)\ge1$. Then $T+1>u$, and
$\log(2/\eps)=(u+4)\log2\le5u\log2$.
Thus $T+1$ is at least a constant times $\log(2/\eps)$. Substituting
$\rho=8\eps$ into the preceding inequality proves the stated lower bound.
\end{proof}

\section{Comparison and remaining questions}
Table~\ref{tab:results} summarizes the general bounds. The lower bound is a
worst-case entry-query bound for sufficiently large matrices.
\begin{table}[ht]
\centering
\renewcommand{\arraystretch}{1.5}
\begin{tabular}{@{}ll@{}}
\toprule
Result & Query complexity\\
\midrule
Upper bound (Theorem~\ref{thm:upper}) &
$O(d^2(d+\log(1/\epsilon))^2/\eps^2)=\tilde O(d^4/\epsilon^2)$\\
Lower bound (Theorem~\ref{thm:lower}) &
$\Omega(d^{3/2}\log(1/\epsilon)/\eps)\ \;  \qquad=\tilde \Omega(d^{3/2}/\epsilon)$\\
\bottomrule
\end{tabular}
\caption{General nonadaptive one-sided testing of binary rank at most $d$.}\label{tab:results}
\end{table}

For constant $\eps$, the bounds leave a gap between $\Omega(d^{3/2})$ and
$O(d^4)$. Three concrete questions remain. First, can one strengthen the
completion argument or choose a different far distribution to increase the
exponent $3/2$ of $d$? Second, can the reconstruction argument reduce the
number of sampled rows and columns, or eliminate the basis-location
logarithm? Third, can a sparse nonadaptive query pattern replace the full
sampled grid while still supplying a one-sided rejection certificate?

The two arguments address complementary obstacles. The upper bound limits
how many low-rank explanations a sample can have. The lower bound shows how
often sparse answers still admit such an explanation. Closing the gap requires
improving at least one of these steps. Throughout, the property is binary
rank \emph{at most} $d$.

\bibliographystyle{plain}
\bibliography{references}

\appendix

\end{document}